\documentclass[useAMS,usenatbib]{mn2e}
\usepackage{newtxtext,newtxmath} % new font
\usepackage{epsfig,comment}
\usepackage{color}
\usepackage{natbib}
\usepackage{deluxetable}
\usepackage{amsmath}
\usepackage{graphicx}
\usepackage[colorlinks=true,citecolor=blue]{hyperref}
\usepackage{epstopdf}

\defcitealias{Rakshit_2020}{R20}
\defcitealias{Calderone2017MNRAS.472.4051C}{C17}
\defcitealias{Shen2011ApJS..194...45S}{S11}

\def\cm2{cm$^{-2}$}

\def\Obh2{\Omega_{\rm b}h^{2}}

\def\J21{J_{21}}

\newcommand{\bit}{\begin{itemize}}
\newcommand{\eit}{\end{itemize}}
\newcommand{\ben}{\begin{enumerate}}
\newcommand{\een}{\end{enumerate}}
\newcommand{\be}{\begin{equation}} 
\newcommand{\en}{\end{equation}}

\def\h2{H$_2$}

\def\kms{kms$^{-1}$}

\twocolumn
\title[Estimating host galaxy fraction from AGN luminosity]{An empirical relation to estimate host galaxy stellar light from AGN spectra}

\author[Jalan Priyanka]{{Priyanka Jalan$^{1,2}$\thanks{E-mail: priyajalan14@gmail.com }, Suvendu Rakshit$^{2}\thanks{E-mail: suvenduat@gmail.com }$, Jong-Jak Woo$^{3}$, Jari Kotilainen$^{4}$, and C. S. Stalin$^{5}$} \\ $^{1}$ Center for Theoretical Physics of the Polish Academy of Sciences, Al. Lotników 32/46, 02-668 Warsaw, Poland \\
$^{2}$ Aryabhatta  Research Institute of Observational Sciences (ARIES), Manora Peak, Nainital, 263002 India \\
$^{3}$ Astronomy Program, Department of Physics and Astronomy, Seoul National University, Seoul 151-742, Republic of Korea \\
$^{4}$ Finnish Centre for Astronomy with ESO (FINCA), FI-20014 University of Turku, Finland\\
$^{5}$ Indian Institute of Astrophysics, Block II, Koramangala, Bangalore-560034, India}
\begin{document}
\date{Accepted 2023 February 2. Received 2023 February 2; in original form 2022 October 3}

\pagerange{\pageref{firstpage}--\pageref{lastpage}} \pubyear{2023}

\maketitle

\label{firstpage}
\begin{abstract}
Measurement of black hole mass for low-$z$ ($z\leq0.8$) Active Galactic Nuclei (AGNs) is difficult due to the strong contribution from host galaxy stellar light necessitating detailed spectral decomposition to estimate the AGN luminosity. Here, we present an empirical relation to estimate host galaxy stellar luminosity from the optical spectra of AGNs at $z\leq 0.8$. The spectral data were selected from the fourteenth data release of the Sloan Digital Sky Survey (SDSS-DR14) quasar catalog having a signal-to-noise ratio at 5100 \AA~(SNR$_{5100}$) $>$10 containing 11415 quasars. The median total luminosity (log ($L_\text{total}$/[erg s$^{-1}$])), stellar luminosity (log ($L_\text{star}$/[erg s$^{-1}$])), and AGN continuum luminosity ((log $L_\text{cont}$/[erg s$^{-1}$])) in our sample are 44.52, 44.06, and 44.30, respectively. We fit the AGN power-law continuum, host galaxy, and iron blend contribution, simultaneously over the entire available spectrum. We found the host galaxy fraction to anti-correlate with continuum luminosity and can be well-represented by a polynomial function, which can be used to correct the stellar light contribution from AGN spectra. We also found anti-correlation between host galaxy fraction and iron strength, Eddington ratio, and redshift. The empirical relation gives comparable results of host-fraction with the image decomposition method.

\end{abstract}

\section{Introduction}
\label{s1:intro}

Active galactic nuclei (AGNs) are powered by the accretion of matter onto a supermassive black hole ($> 10^5 M_\odot$) with luminosity ranging from $10^{44}-10^{48}$ erg s$^{-1}$. Previous studies suggest that nearly all massive galaxies host supermassive black holes (SMBHs) and the properties of these SMBHs show correlations with the properties of the host galaxy, indicating a strong connection between them \citep[see ][for a review]{Kormendy2013ARA&A..51..511K}. Massive black holes may originate before the era that marks the peak of galaxy formation (at $z\sim2$), as evident by the history of star formation in bright galaxies \citep[][]{Madau2014ARA&A..52..415M}. As a result, it gives credence to theoretical arguments that spheroid formation and the growth of SMBHs are closely linked and the associated extensive energetic outflows from AGN may have a significant correlation with physical processes in the host galaxy. 

To understand this relation between black holes and their host galaxy, it is crucial to emphasize their spectral decomposition \citep[][]{Greene_2005, Shen2011ApJS..194...45S, Bongiorno2014MNRAS.443.2077B, Varisco2018A&A...618A.127V, Rakshit_2020}. This is especially important at low-$z$ in optical wavelengths where the contamination by their host galaxy is significant. The advent of large surveys such as the Sloan-Digital Sky Survey (SDSS) and Dark Energy Spectroscopic Instrument (DESI), which capture millions of such AGNs,  motivates us to find an empirical relation between fraction of luminosity from black hole and their host galaxy, that can be used for further studies. For example, black hole masses in AGNs are mostly calculated using the virial relation given by the reverberation mapping study, which uses the luminosity of the continuum to find the size of the broad line region (or radius-luminosity relation) \citep[e.g.,][]{ Shen2011ApJS..194...45S, Rakshit_2020}. Non-removal of the host galaxy may lead to an overestimation of the luminosity and hence the black hole mass. 
\par Various methods have been developed to decompose the host galaxy contribution correctly and to maximize the use of archival data. Recent studies by \citet[][hereafter \citetalias{Rakshit_2020}]{Rakshit_2020} used SDSS-DR14 quasar sample and decomposed host galaxy and quasar contribution using principal component analysis (PCA). Similarly, \citet[][hereafter \citetalias{Calderone2017MNRAS.472.4051C}]{Calderone2017MNRAS.472.4051C} and \citet[][]{Varisco2018A&A...618A.127V} used SDSS-DR10 sample at $z<0.6$ and $z<0.8$, respectively to account for this host galaxy contamination. \citetalias{Calderone2017MNRAS.472.4051C} developed an automated software package, QSFIT\footnote{https://github.com/gcalderone/qsfit} that decomposes the spectrum into the host galaxy and AGN continuum, the Balmer continuum emission, optical and ultraviolet iron blended complex, and emission lines. They have used a power-law model for AGN and an elliptical galaxy template for the host galaxy. Similarly,  \citet[][]{Shen2011ApJS..194...45S} estimated the host galaxy contamination for $z\leq0.5$ quasars using stacked spectra assuming that the highest luminosity bin (log ($L_\text{cont}$/[erg s$^{-1}$]) = 45.5) is not affected by host galaxy contamination. \citet[][]{Vanden_Berk_2006} separated the host galaxy using the PCA method from the broad line AGN for 4666 spectra from SDSS at $z\leq 0.75$ and found dependence on various parameters such as signal-to-noise ratio (SNR) and host galaxy class.
\par
Detailed spectral decomposition is necessary to remove the host contamination, however, high S/N spectra along with clear detection of absorption lines are required for such spectral decomposition. Unfortunately, a majority of the spectra obtained in large surveys do not have high S/N.
In this paper, we carried out a detailed spectral decomposition for $z<0.8$ AGNs in the SDSS-DR14 \citep[][]{SDSS2018A&A...613A..51P} catalog by performing an independent fit of each AGN spectrum to obtain an empirical relation between host galaxy fraction and AGN continuum luminosity that can be used to estimate the host-fraction for low-S/N spectra where detailed spectral decomposition is difficult to perform. The paper is organized as follows: in Sect.~\ref{s:data}, we discuss the spectral data used and its various components. In Sect.~\ref{s3:result}, we detailed our results followed by a comparison with previous studies. We also analyzed the variation of host galaxy fraction with redshift, SNR and other AGN parameters. Finally, we conclude our work in Sect.~\ref{s:conc}.
Throughout, we used a flat background cosmology with $\Omega_m$ = 0.3, $\Omega_\lambda$ = 0.7, and $H_o$ = 70 \kms Mpc $^{-1}$.

\section{Data and spectral fitting}
\label{s:data}
We used the complete compilation of the SDSS-DR14 quasar catalog and more sophisticated parameter measurements from \citetalias{Rakshit_2020}. We limited the analysis to $z\leq0.8$ sources for which SDSS covers $\lambda < 5000$ \AA~(rest-frame) providing a better constraint to the host galaxy contribution, resulting in 55427 sources. The host galaxy fraction (hf $\equiv L_\text{star}/L_\text{total}$) can not be precisely measured  for a very low-SNR spectrum therefore we limited the sample with  SNR at 5100 \AA~(SNR$_{5100})>10$. This includes 11415 spectroscopically confirmed quasars with a median SNR$_{5100}$ of $\sim 15$.

\begin{figure}
\centering
\includegraphics[height=5cm,width=9.0cm]{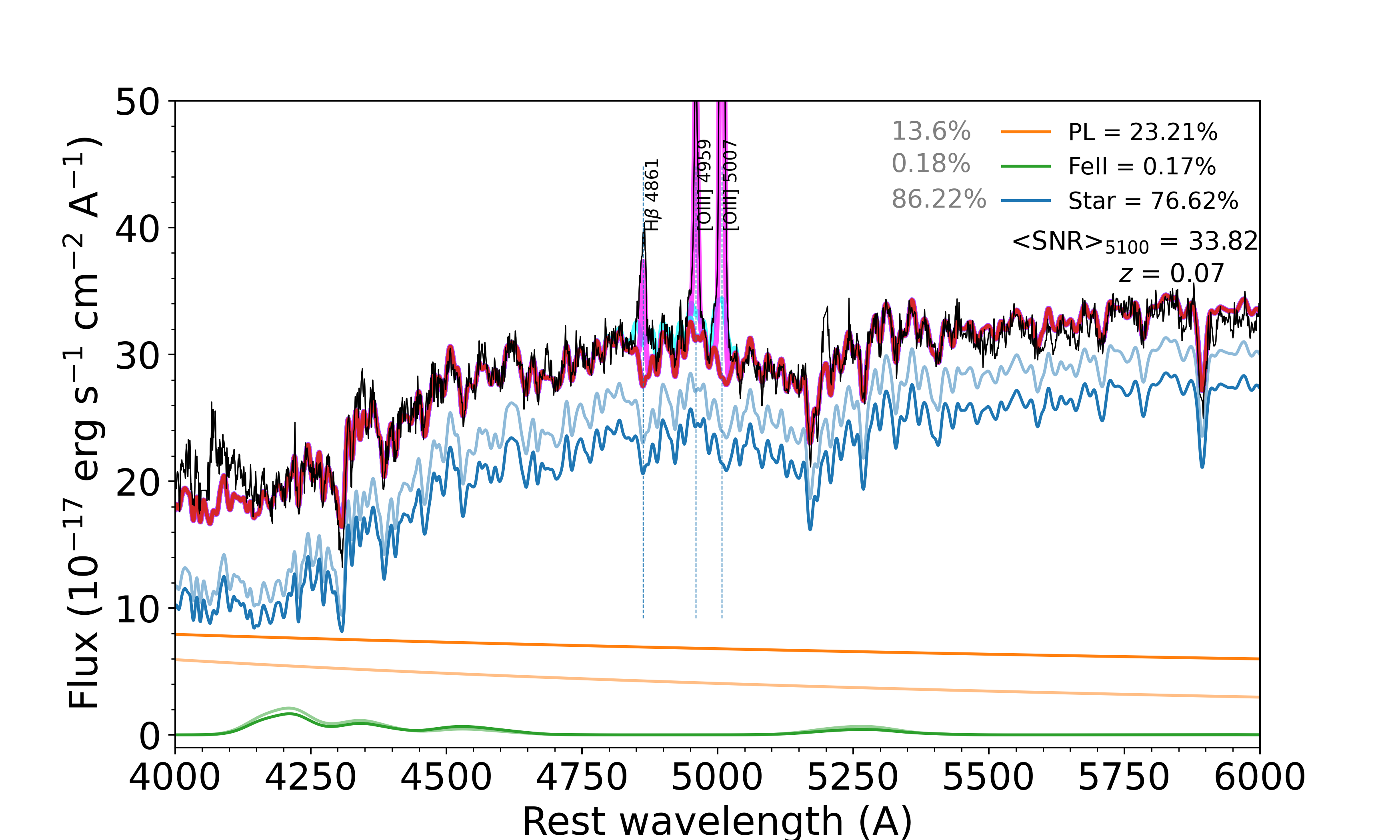}
\includegraphics[height=5cm,width=9.0cm]{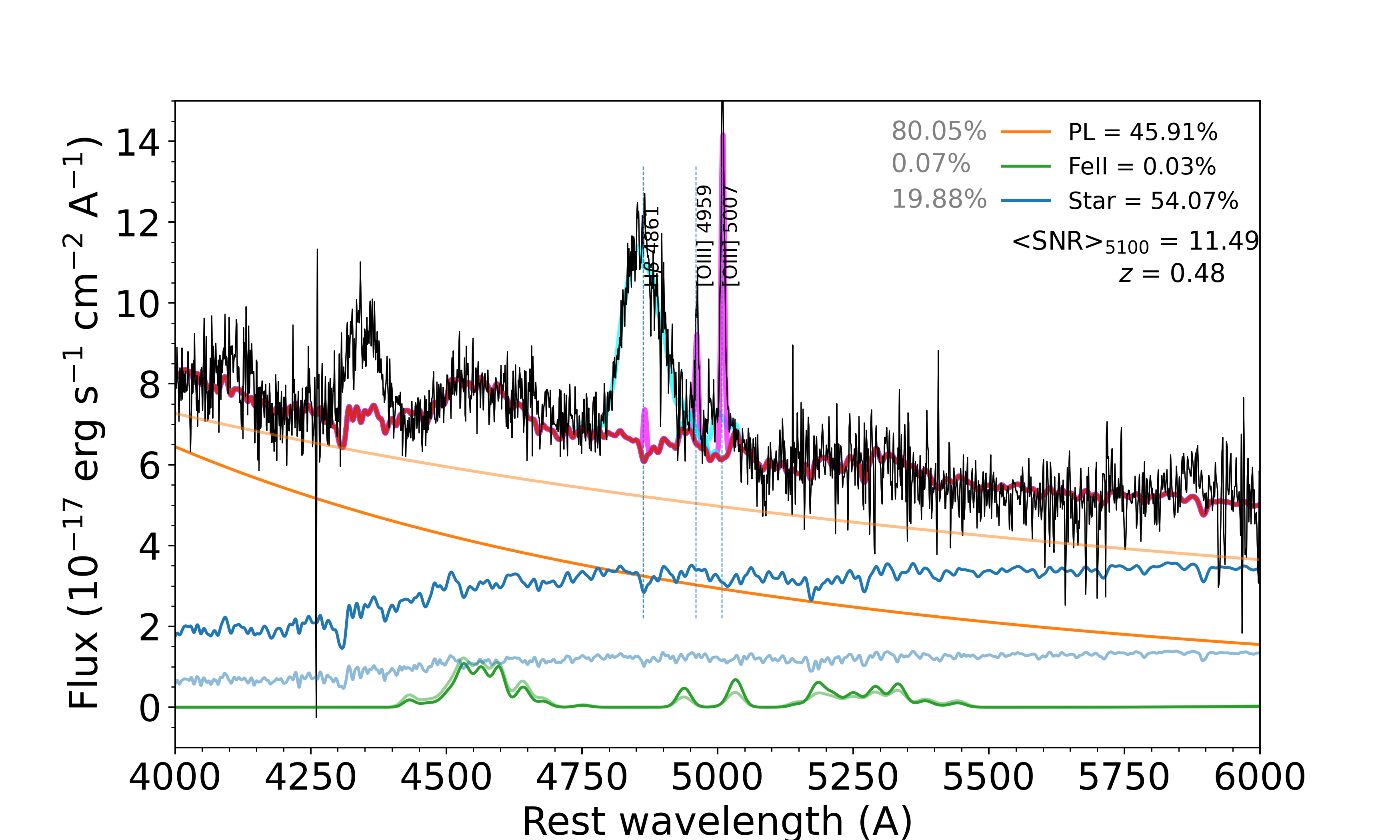}
\caption {The plot shows two spectra with their individual components. The AGN continuum, iron continuum and host galaxy fraction are shown in orange, green and blue colour, with contributions mentioned as PL, Fe~{\sc ii} and star, respectively. The fainter colour shows the spectral decomposition for the constant spectral slope of -1.7 (see text in Sect.~\ref{s4:compare}). The top panel shows an example of high galaxy contribution with higher SNR $\sim34$ at $z\sim0.07$ and the bottom panel shows an example of $\sim 54\%$ galaxy contribution at lower SNR $\sim 11$ at $z\sim0.48$. The plot also shows broad (cyan) and narrow (fuchsia) components of the emission lines.}
\label{Fig:fig1}
\end{figure}

Each observed spectrum was corrected for Galactic extinction using the \citet[][]{Schlegel1998ApJ...500..525S} extinction map and the Milky Way extinction law with $R_V=3.1$ from \citet[][]{Cardelli1989ApJ...345..245C}.
Then the spectra were converted to their respective rest frame (based on their redshift). After masking the strong emission lines, we modeled each spectrum using a combination of starlight from the host galaxy, AGN power-law and iron blends with their contribution varying from source to source, as discussed below.\\
{\bf Star light}: The photons from the stars present in the host galaxy contribute to the total AGN spectrum, especially at low$-z$ in optical wavelengths. In order to model the stellar contribution, we used the stellar template from the  Indo-US  spectral  library \citep[][]{Valdesarticle}, which has been previously used by various authors \citep[][]{Rakshit_2018}. It contains seven spectra of G- and K-type giant stars of various temperatures with a resolution of 1.35\AA~ and covering the wavelength range of 3650-9200 \AA. The templates were broadened by a Gaussian kernel with line width as a free parameter along with the velocity shift as another free parameter. The weights were optimized for each stellar template.\\ 
{\bf AGN continuum:}
The AGN contribution is simply a power law of the form
\begin{equation}
  L_\lambda =   A \times (\lambda/5100)^B
  \label{eq:agn_slope}
\end{equation}
where A is the luminosity density at 5100 \AA~or the normalization parameter and B is the spectral slope. The line-free region of the entire spectral range is used for this continuum fitting. Even for large spectral coverage, we assumed a single power-law however, a broken power-law can also be modeled. However, in the sources that have star-light contributions similar to or higher than the AGN contribution the slope degenerates with the star-light and hence the broken power law cannot be well constrained.\\ 
{\bf Iron lines:} 
We considered the spectral range of 4000-6000 \AA\, to perform the spectral fitting and modeled the optical Fe~{\sc ii} emission using the Fe~{\sc ii} templates constructed by \citet[][]{Kovavcevic2010ApJS..189...15K}.   Previous studies found that this template provides a better fit to the Fe~{\sc ii} emission in AGN \citep[][]{Barth2015ApJS..217...26B}. The UV component of the Fe~{\sc ii} emission ranging from 1250-3090 \AA~is not relevant in the present decomposition i.e., $>3650$ \AA. The \citet[][]{Kovavcevic2010ApJS..189...15K} Fe~{\sc ii} template consists of five different templates representing different groups of multiplet, therefore, normalization of each template was needed. The Fe~{\sc ii} template spectra were convolved using a Gaussian broadening kernel as done for the stellar template. The width and the shift of all five templates were kept the same for a given spectral fit. The total continuum is a combination of the host galaxy, AGN power-law and Fe~{\sc ii} emission.\\
{\bf Emission lines:} After the subtraction of the total continuum from the spectrum the emission lines (broad and narrow) were fitted. The emission line complex  is fitted over the spectral range of 4600-5050 \AA.
We fitted the broad component using a 6-th order Gaussian Hermite function and each narrow component using a single Gaussian. The O~{\sc iii} doublets were fitted using double Gaussian. Using the best-fit model, we extracted the parameters such as emission line flux and FWHM which is used to calculate the black hole mass and Eddington ratio \citep[using][]{Rakshit_2018}. 
\par The best-fit model was found by performing a nonlinear Levenberg–Marquardt least-squares minimization using MPFIT3 \citep[][]{Markwardt2009ASPC..411..251M} code in IDL. This allowed us to properly decompose different model components estimating host galaxy fraction and the host-subtracted continuum luminosity at 5100 \AA. 
Fig.~\ref{Fig:fig1} presents an overview of the results of our continuum fitting procedures. The top and bottom panel shows a spectrum with high SNR ($\sim 34$) at $z=0.07$ and SNR ($\sim 11$) at $z=0.48$, with $\sim 76\%$ and $\sim 54\%$ host contribution, respectively. Individual decomposed components are also shown.  

\begin{figure}	
\centering
\includegraphics[height=4.5cm,width=4cm]{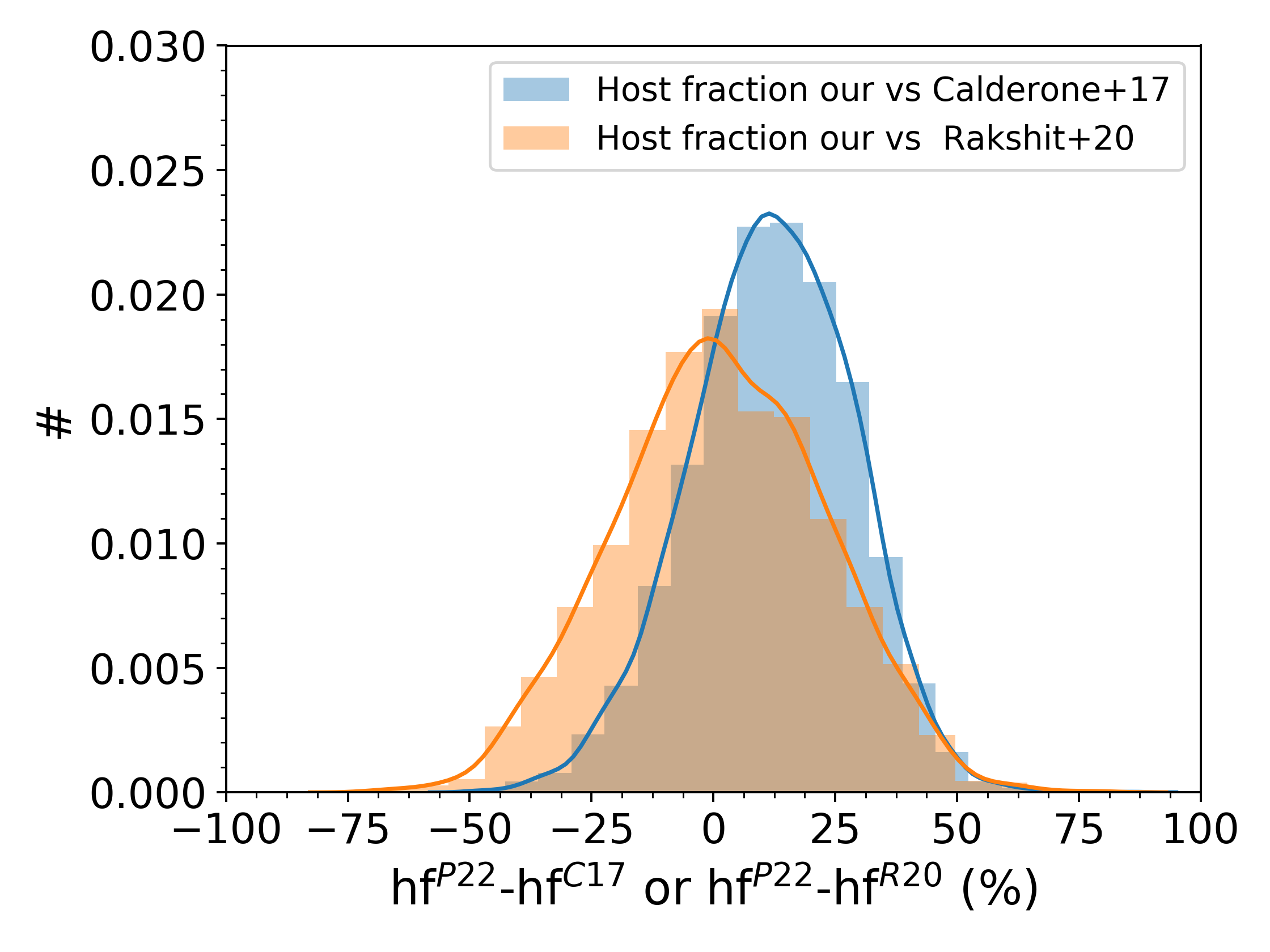}
\includegraphics[height=4.5cm,width=4cm]{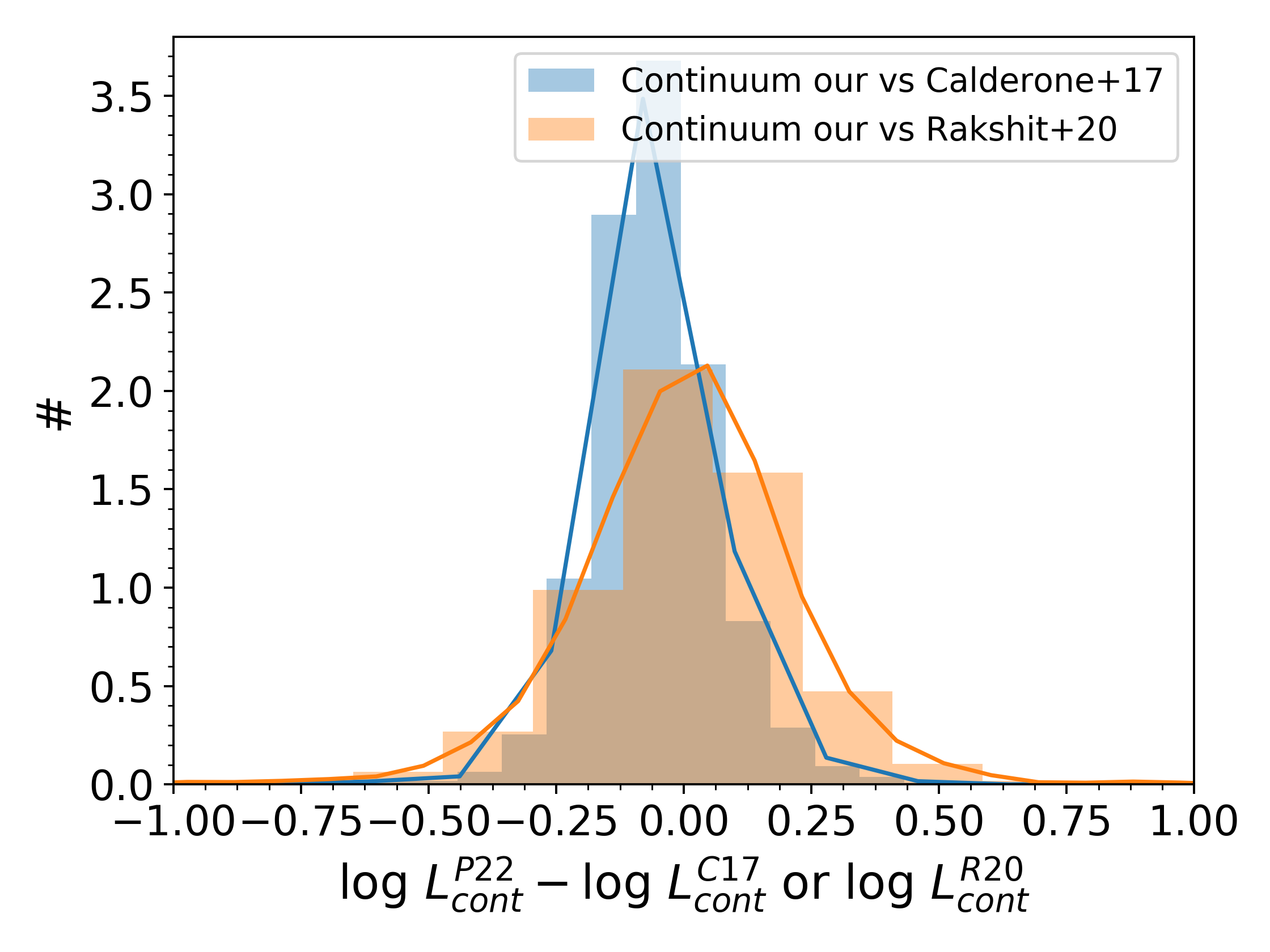}
\caption{Comparing host galaxy fraction (left) and continuum luminosity (right) calculated in this work (P22) with \citetalias{Calderone2017MNRAS.472.4051C} and \citetalias{Rakshit_2020} at 5100 \AA. Histograms of their differences are shown.}
\label{Fig:fig_compare}
\end{figure}

  \begin{figure}	
    \centering
  \includegraphics[height=6.2cm,width=7cm]{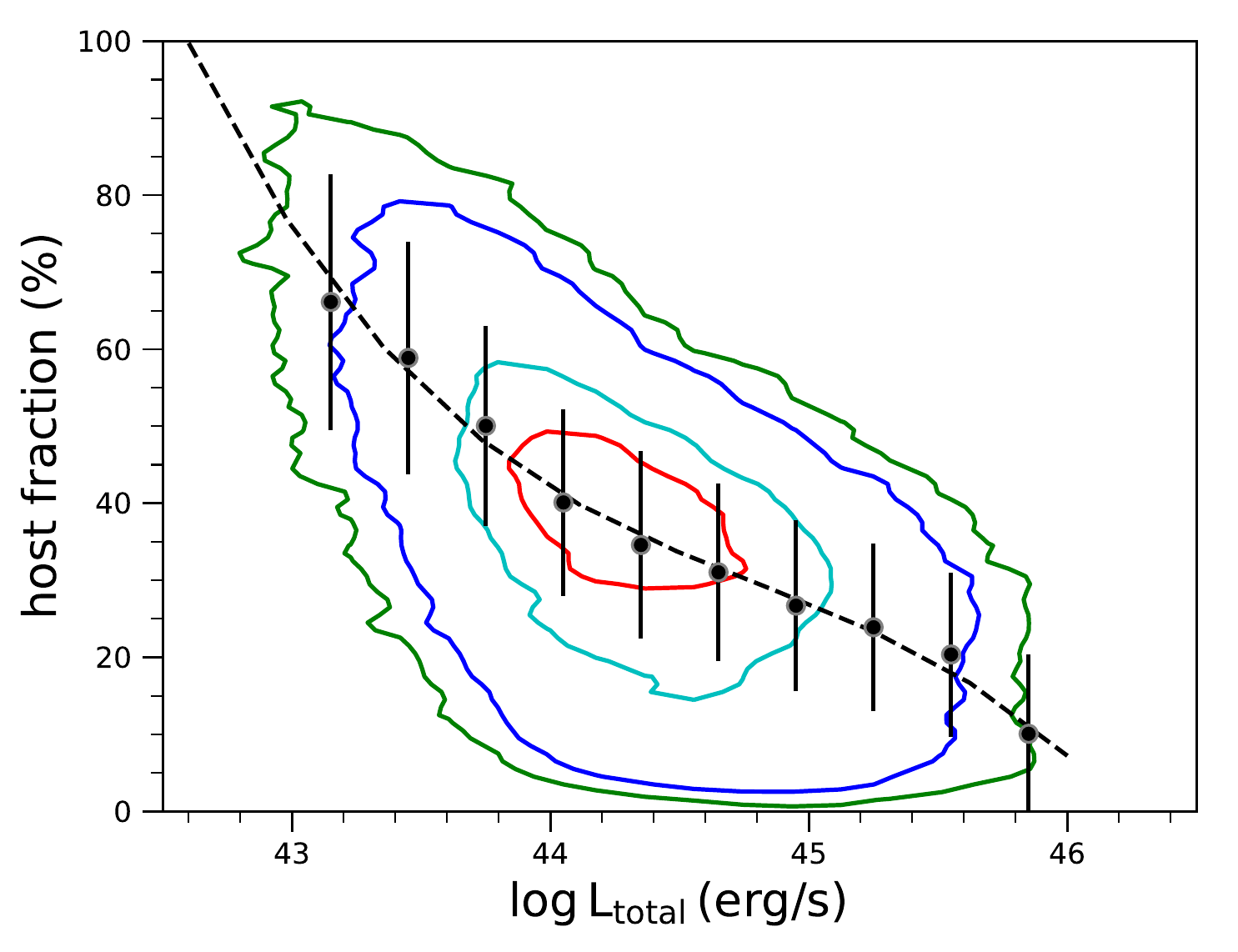}
  \includegraphics[height=6.2cm,width=7cm]{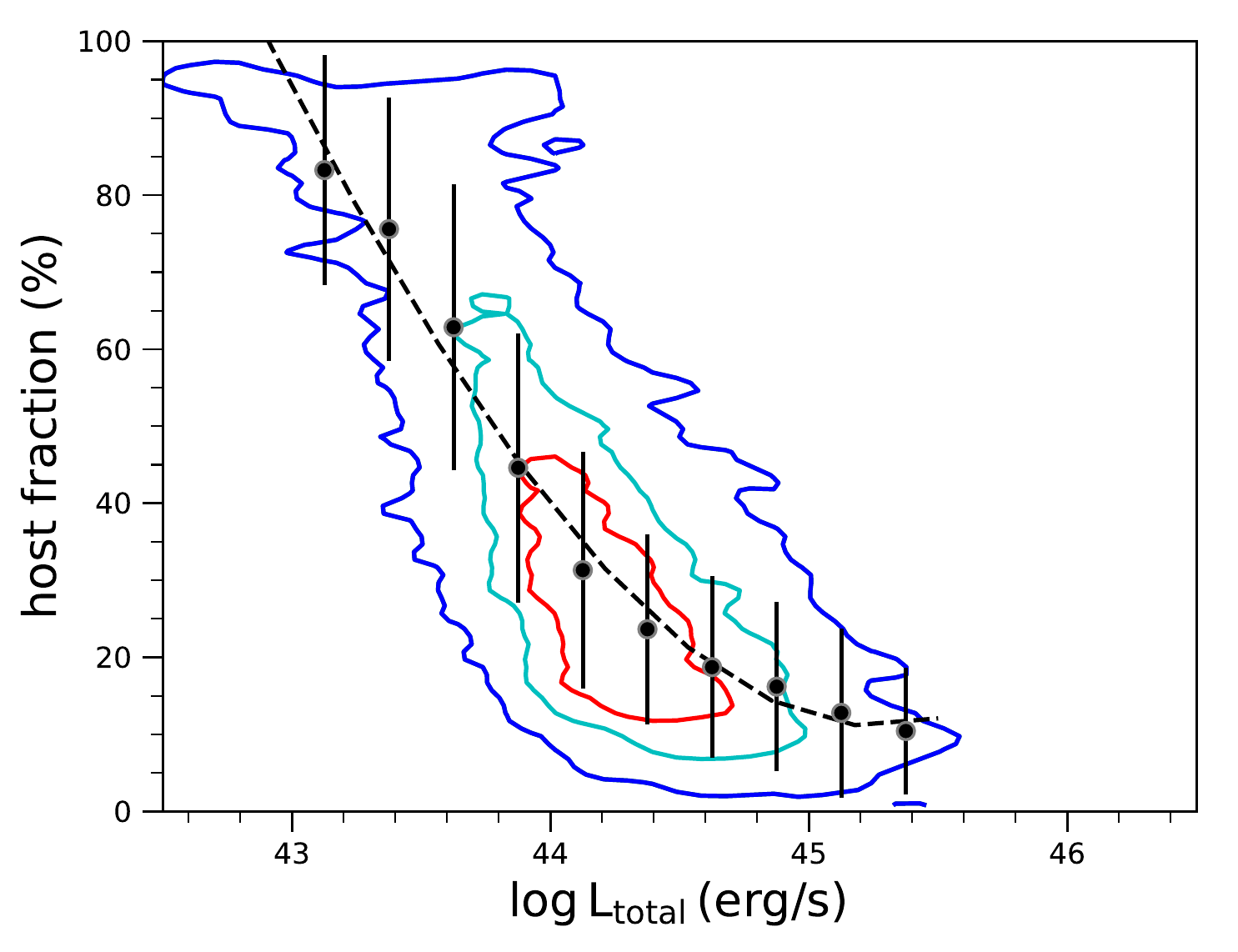}
  \caption{{\it Top panel: } The density contours represent the host luminosity fraction to that of total luminosity at 5100 \AA~versus  continuum luminosity at 5100 \AA~in this work. The 40 percentile, 1$\sigma$, 2$\sigma$ and 3$\sigma$ contours are shown. The points represent the median of the sample in each luminosity bin with error-bars showing the standard deviation in the bin. The best-fit polynomial fit of the binned data is shown by the dashed line. {\it Bottom panel: } Same as the top panel but for spectral slope fixed to -1.7.}
  \label{Fig:fig_main}
  \end{figure}

  \section{Results}
\label{s3:result}

\subsection{Comparison with previous studies}
\label{s4:compare}
We compare our results of host galaxy fraction estimate with that obtained by \citetalias{Calderone2017MNRAS.472.4051C} and \citetalias{Rakshit_2020}, enlisting the similarities and dissimilarities. \citetalias{Rakshit_2020} first decomposed the spectrum using 5 PCA components for the galaxy and 20 PCA components for quasars. After subtracting the galaxy contribution, they fitted the residual with a power-law continuum and iron template masking the emission lines. On the other hand, \citetalias{Calderone2017MNRAS.472.4051C} fitted different components of the continuum  and emission lines simultaneously where the host galaxy is represented by a single elliptical galaxy template of \citet[][]{Mannucci2001MNRAS.326..745M} with the only free parameter being the normalization factor. \par
For the AGN contribution, \citetalias{Rakshit_2020} uses a similar power-law as mentioned in Eq.~\ref{eq:agn_slope}. However, unlike our varying spectral slope, \citetalias{Calderone2017MNRAS.472.4051C} assumes a fixed  value  of $B = -1.7$ for low-redshift ($z \leq 0.6$) AGNs due to the difficulty of separating stellar and non-stellar components in AGNs closer than $z \sim 0.7$. Similar to \citetalias{Calderone2017MNRAS.472.4051C}, we also tested our spectral decomposition for the spectra with a constant spectral slope. We show this spectral decomposition in fainter colour in Fig.~\ref{Fig:fig1}. \par 
We compare the host galaxy fraction percentage calculated in this work with that obtained by \citetalias{Calderone2017MNRAS.472.4051C} and \citetalias{Rakshit_2020} for the common sources in Fig.~\ref{Fig:fig_compare}.
The mean difference of the host galaxy fraction between this work and \citetalias{Rakshit_2020} is 1.51$\pm$21.34\% while this work and \citetalias{Calderone2017MNRAS.472.4051C} is 11.92$\pm$16.64\%. This suggests our host galaxy fraction is more consistent with \citetalias{Rakshit_2020} than \citetalias{Calderone2017MNRAS.472.4051C} measurements although with a larger scatter with \citetalias{Rakshit_2020} compared to \citetalias{Calderone2017MNRAS.472.4051C}. A large fraction of sources with host galaxy fraction $\sim$30\% as estimated in this work have almost nil host contamination in \citetalias{Calderone2017MNRAS.472.4051C}. The difference in the host contribution measurement by different authors could be mainly due to the use of different host galaxy decomposition methods. Since \citetalias{Calderone2017MNRAS.472.4051C} assumed the host galaxy to be an elliptical galaxy and low-luminous AGNs are found to be hosted in spiral galaxies, such an assumption could underestimate the host galaxy flux measurement, especially in low-luminosity AGNs \citep[][and references therein]{Crenshaw_2003, Olguin10.1093/mnras/stz3549,Gkini2021A&A...650A..75G}. The right panel of Fig.~\ref{Fig:fig_compare} shows the consistency of the continuum luminosity with \citetalias{Rakshit_2020} and a slight difference with respect to estimates by \citetalias{Calderone2017MNRAS.472.4051C}.
Furthermore, we checked if using a fixed spectral slope by \citetalias{Calderone2017MNRAS.472.4051C} could be the reason and we found the mean difference of the host galaxy fraction of this work and \citetalias{Calderone2017MNRAS.472.4051C} reduces to 6.37$\pm$23.77\%, if the spectral slope fixed to $B=-1.7$. On the other hand, \citetalias{Rakshit_2020} uses the PCA method which assumes that each spectrum can be defined by combining two independent sets of eigen-spectra from the pure galaxy and the pure quasar samples. Moreover, only if the host galaxy fraction in the wavelength range of 4160-4210 \AA~is larger than 10\% the decomposition is considered effective whereas we apply no such constraint.

\subsection{Host galaxy fraction vs. continuum luminosity}
The correlation between the host galaxy fraction and the luminosity at 5100 \AA~ is shown in Fig.~\ref{Fig:fig_main}. The median 5100 \AA~total luminosity (log ($L_\text{total}$/[erg s$^{-1}$])), stellar luminosity (log ($L_\text{star}$/[erg s$^{-1}$])) and  AGN continuum luminosity (log ($L_\text{cont}$/[erg s$^{-1}$])) in our sample are  44.524, 44.062 and 44.301, respectively. All the luminosity (i.e, $L_\text{star}$, $L_\text{cont}$ and $L_\text{total}$) estimates are found to be increasing with the redshift, with varying slopes. This is contributed by two factors: (i) there is a bias in selecting more luminous quasars at higher$-z$ due to flux-limited observations and (ii) more host galaxy light will be available in the 3$^{\prime\prime}$ SDSS aperture for higher$-z$ targets because of the smaller angular size of the host galaxy.
However, the continuum luminosity is found to have a steeper slope with redshift as compared to the stellar luminosity.

It is evident from Fig.~\ref{Fig:fig_main} that the host galaxy luminosity fraction decreases as the continuum luminosity increases. This is clear from the (median) binned data plot which shows a decreasing trend.
The host galaxy luminosity fraction is dominant with $>$50\% for the continuum luminosity $<10^{43.7}$ erg s$^{-1}$. As expected, at a given $L_\text{cont}$, there is a large dispersion in the host galaxy fraction. This is because, for a given $L_\text{cont}$ there would be sources with a distribution of Eddington ratios, which means sources with a range of black hole mass (or a range of $L_{\mathrm{star}}$ due to the correlation between black hole mass and stellar mass) leading to a large distribution of host-fraction.

In order to find a mathematical relation between the host galaxy fraction with the total AGN luminosity, we fitted the binned data in the top panel of Fig.~\ref{Fig:fig_main} using a third-order polynomial and found
\begin{equation}
y = (42.01\pm1.15) + (-20.53\pm 1.47) x +  (9.34\pm2.45) x^2 +  (-3.88\pm 1.33) x^3
\label{eq:main}
\end{equation}
where $x$ = $\log$ L$_{\mathrm{total}}$ - 44.

We estimate the host-fraction of individual objects based on the above empirical relation and compare them with the direct measurement from model fitting. The mean difference of host-fraction estimated between the two methods is $\sim$0\% with a dispersion of $\sim$15\%. As mentioned above, we also calculated the host galaxy fraction by fixing the spectral slope to be -1.7, and the modified host galaxy fraction versus total luminosity is shown in the bottom panel of Fig.~\ref{Fig:fig_main}. The modified equation for fixed spectral slope is 
\begin{equation}
y = (39.87\pm1.96) + (-42.52\pm 3.08) x +  (13.39\pm3.93) x^2 +  (1.73\pm 3.52) x^3.
\end{equation}

 \begin{figure}	
   \centering
   \includegraphics[height=3.5cm,width=4cm]{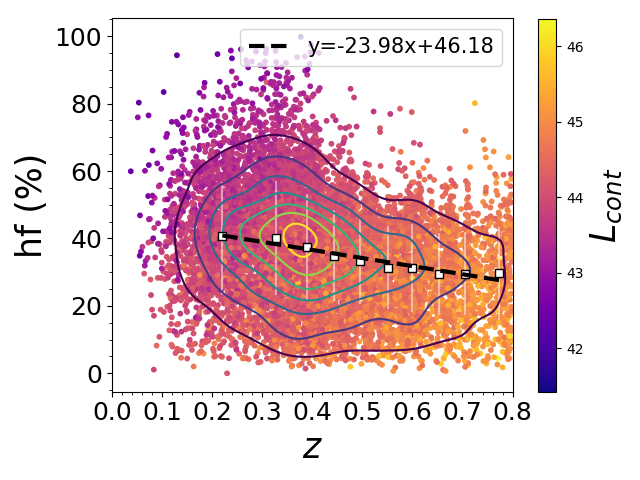} 
    \includegraphics[height=3.5cm,width=4cm]{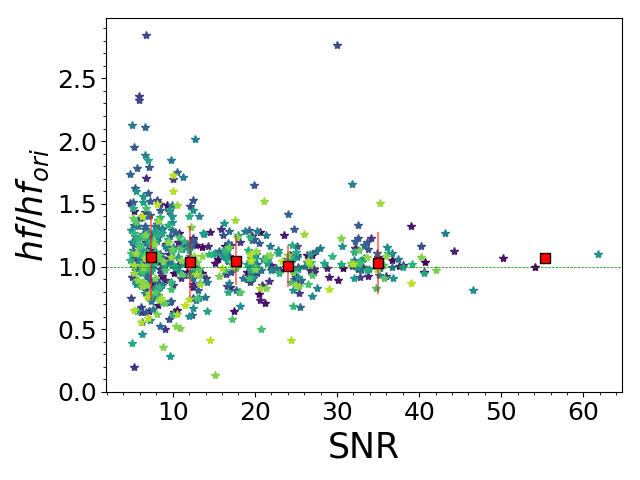}
\caption {Fraction of host luminosity as compared to the total luminosity at $\lambda \sim 5100$ \AA~as a function of redshift [left panel] and SNR at 5100 \AA~[right panel]. Multiple objects are shown with stars in different colours. The red squares represent the median in each bin.}
  \label{Fig:fig_z_snr}
 \end{figure}
 
 \begin{figure}	
   \centering
   \includegraphics[height=3.cm,width=4.cm]{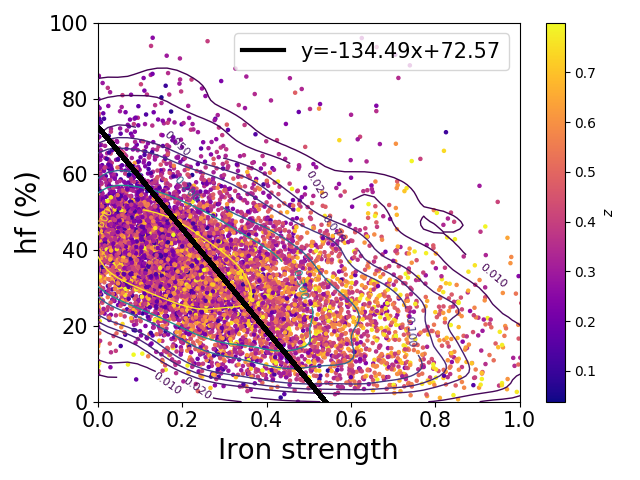}
   \includegraphics[height=3.cm,width=4.cm]{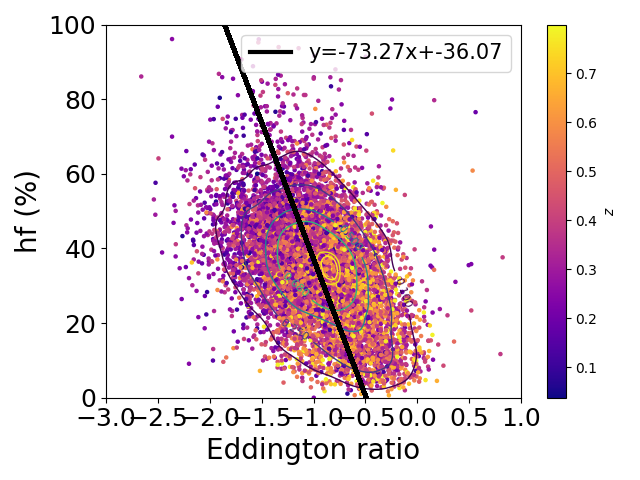}
    \caption{The plot shows host luminosity as compared to the total luminosity at $\lambda \sim 5100$ \AA~versus iron strength [left panel] and Eddington ratio [right panel], respectively. The colour bar represents the redshift. The black line represents the robust linear regression fitting using BCES.}
  \label{Fig:fig_param}
 \end{figure}

\subsection{Host galaxy fraction with redshift and SNR}
\label{s5:zandsnr}
In Fig.~\ref{Fig:fig_z_snr} (left panel), we show the  host galaxy fraction as a function of redshift. We found that the host galaxy fraction percentage reduces as we go to higher redshifts, which is expected. 
With the increasing redshift, the continuum luminosity is found to increase (as shown by the colour bar), and thus the host fraction decreases. 
Spectral decomposition is sensitive to the SNR, and higher reliability of decomposition is expected for higher SNR spectra.  In order to study the dependence of SNR on the host galaxy fraction, we used a subsample of high SNR data which have a host galaxy fraction (hf$_\text{ori}$) $>20$\%. We calculated the host galaxy fraction (hf) for spectra after degrading their SNR. For this, we first multiplied a factor of 2 to 10 by the original flux error for each high SNR spectra. We then added to the original flux spectra a Gaussian random deviation of zero mean and standard deviation given by the new flux errors. We then performed the spectral decomposition from these mock spectra as done for the original spectra and calculated the host galaxy fraction.  We evaluated the ratio of hf/hf$_\text{ori}$ as plotted in the right panel of Fig.~\ref{Fig:fig_z_snr}, here the red squares show the median values. As evident from the plot, the scatter at lower SNR $<10$ is larger although the median value is consistent with unity. Therefore, the host galaxy fraction estimates are reliable only for SNR $>10$. This justifies the use of spectra having SNR $>10$ for this work. 
 
\subsection{Host galaxy fraction and AGN parameters}
\label{s6:agn}

\par To study the dependency of host luminosity fraction with AGN parameters, we estimated black hole mass and Eddington ratio. The black hole mass was calculated using the virial relation from the single-epoch spectrum based on the FWHM of the H$\beta$ emission line and AGN luminosity at 5100A using the relation given by \citet{2015ApJ...801...38W}. The Eddington ratio was estimated by the ratio of Bolometric to Eddington luminosity, where the latter was calculated from the black hole mass \citep{Rakshit_2018}. Previous studies found an anticorrelation between iron strength and [O III] while a strong positive correlation between iron strength and Eddington ratio \citep[e.g.,][]{1992ApJS...80..109B,Rakshit_2017}. To study the correlation between host-fraction and iron strength, we estimated iron strength using the ratio of optical Fe~{\sc ii} flux of the Fe~{\sc ii} multiplets integrated over the wavelength range of 4434–4684 \AA~to H$\beta$ broad component. In the left panel of Fig.~\ref{Fig:fig_param}, we show the host galaxy fraction versus the iron strength. We found host galaxy fraction is anti-correlated with the iron strength. Since high accreting sources are strong Fe II emitters, therefore, the host galaxy fraction should be anti-correlated with the Eddington ratio also. As expected, the  host galaxy fraction reduces with the increment in the Eddington ratio as shown in the right panel of Fig.~\ref{Fig:fig_param}.

\subsection{Impact of limited stellar templates}
\label{s:caveat}
In this study, we have not included OB stellar templates in our stellar template list (e.g., see Sect.~\ref{s:data}) due to their strong blue (along with the lack of absorption features in the wavelength range of 4000-6000 \AA) spectra that is indistinguishable from the AGN power-law continuum. This leads to a high degeneracy in the model decomposition. Therefore, we can expect a systemic underestimation of the host fraction, especially, for AGNs with a high fraction of a young stellar population. To test the impact of including OB star templates, we generated $\sim200$ spectra by adding OB star templates. We added 50\% host fraction (at 5100 \AA) of the total continuum using the OB star template to the spectrum of high luminous AGNs that originally had almost negligible host galaxy and iron contributions. We fitted the simulated spectrum with stellar templates, including and excluding OB star templates from the stellar template list, and found the mean difference to be 16.94$\pm$6.62\% in this case. This implies that for the young population, our empirical relation may underestimate host galaxy fraction by upto $\sim17\%$. We also performed the simulations for two extreme cases by adding 1) a host fraction of 10\%  that resulted in the host fraction being underestimated by 10.00$\pm$9.84\% and 2) a host fraction of 80\% which gives much higher uncertainty with the host fraction underestimated by 37.75$\pm$3.48\%, if OB star templates are excluded.

\subsection{Applications of the empirical relation}
\label{s:application}
\begin{figure}	
\centering
\includegraphics[height=5.cm,width=5.5cm]{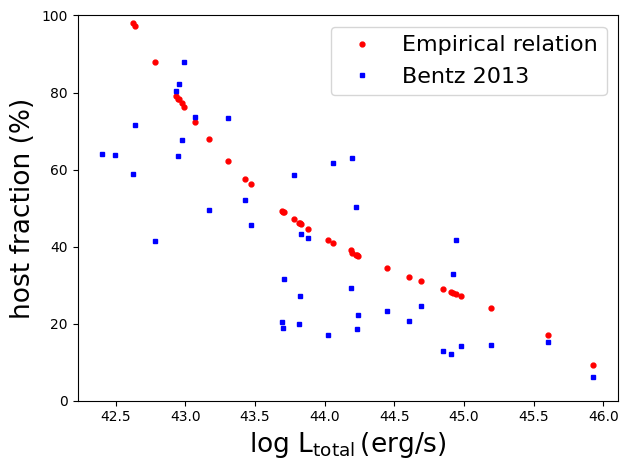}
\caption{The plot shows the host fraction values calculated by \citet[][]{Bentz_2013} (blue squares) using the surface brightness decomposition and using the empirical relation derived in this work in Eq.~\ref{eq:main} (red circle) as a function of total luminosity.}
\label{Fig:figbentz}
\end{figure}

We have investigated the applicability of the empirical relation derived in this work (Eq.~\ref{eq:main}), to a completely independent sample having host galaxy measurement from an independent method. We used the sample from \citet[][hereafter, B13]{Bentz_2013}. Unlike here, B13 used surface brightness decomposition on the Hubble Space Telescope (HST) galaxy images and remove the AGN contribution to find the host galaxy from which the starlight contribution could be measured. They used GALFIT \citep[][]{Peng_2002}, a nonlinear least-squares two-dimensional image fitting algorithm. The host galaxy fraction calculated in B13 is shown with blue-squares in Fig.~\ref{Fig:figbentz}. We use the total luminosity values provided in Table~12 and 13 of B13 and calculate the host galaxy fraction using our empirical relation (Eq.~\ref{eq:main}) as shown in red-circles in Fig.~\ref{Fig:figbentz}. We compare these host galaxy fractions and find the mean difference to be 10.5$\pm$17.3\%. This implies that the empirical relation derived in Eq.~\ref{eq:main}, provides acceptable results even for a sample not observed with SDSS.

\section{Summary}
\label{s:conc}
With the advent of large ongoing/upcoming surveys, such as DESI and SDSS, it is important to quantify host galaxy contribution to the AGN spectra in order to reliably estimate black hole mass. In this work, we provide an empirical relation of the host galaxy's stellar luminosity fraction to that of the total luminosity as a function of the total luminosity. For this, we decomposed the host galaxy contribution from  AGNs at $z\leq0.8$ using the optical spectra from SDSS-DR14 with SNR$_{5100} >10$ leading to 11415 SDSS spectra and estimated host galaxy fraction by fitting stellar light, AGN power-law continuum and iron blends, simultaneously. The empirical relation obtained from our spectral decomposition method is found to be well-reproducing host-fraction for an independent sample of AGNs where host-fraction was estimated using the image decomposition method.
By simulating the spectra at various SNRs, we found the estimated host galaxy fraction to be more reliable for SNR $>10$. We find that the host galaxy fraction is anti-correlated with the total luminosity and can be well-represented with a 3rd-order polynomial function that can be used to correct the spectrum for host galaxy stellar light contribution to estimate pure AGN continuum luminosity. We found the host galaxy fraction is $\geq$ 50\% for total luminosity of (log L$_\text{total}$/[erg s$^{-1}$]) $\leq$ 43.7. A negligible reduction of host galaxy fraction is observed with increasing redshift. The host galaxy fraction is found to be anti-correlated with iron strength and Eddington ratio. 

\section*{Acknowledgement}
 We thank the anonymous referee for his/her valuable comments and constructive suggestions which have greatly improved the manuscript. PJ was supported by the Polish National Science Center through grant no. 2020/38/E/ST9/00395. SR acknowledges the partial support of SRG-SERB, DST, New Delhi through grant no. SRG/2021/001334.

\vspace{-0.6cm}
\section*{Data availability}
The data underlying this article were accessed from SDSS data achieve (\url{https://www.sdss.org/}), managed by the Astrophysical Re-
search Consortium for the Participating Institutions of the SDSS-III
Collaborations. The derived data generated in this research will be shared upon reasonable request to the corresponding author.

\bibliography{references}

\label{lastpage}
\end{document}